\documentclass[%
 reprint,twocolumn,
 amsmath,amssymb,
 aps,
]{revtex4-1}
\providecommand{\python}{{\sc Python }}
\usepackage{graphicx}
\usepackage{dcolumn}
\usepackage{bm}

\begin{document}

\title{Indifferent electromagnetic modes: bound states and topology}

\author{S. A. R. Horsley}
\email{s.horsley@exeter.ac.uk}
\affiliation{Department of Physics and Astronomy, Stocker Road, University of Exeter, Exeter EX4 4QL}


\begin{abstract}
At zero energy the Dirac equation has interesting behaviour.  The asymmetry in the number of spin up and spin down modes is determined by the topology of both space and the gauge field in which the system sits.  An analogous phenomenon also occurs in electromagnetism.  Writing Maxwell's equations in a Dirac--like form, we identify cases where a material parameter plays the role of `energy'.  At zero `energy' we thus find electromagnetic modes that are indifferent to local changes in the material parameters, depending only on their asymptotic values at infinity.  We give several examples, and show that this theory has implications for non--Hermitian media, where it can be used to construct permittivity profiles that are either reflectionless, or act as coherent perfect absorbers, or lasers.
\end{abstract}

\maketitle
\par
Topology is not often applied to physics.  Most physical theories are concerned with \emph{local} behaviour, and topology is about invariant \emph{global} properties.  Nevertheless there are some fascinating examples; characterising Skyrmions in magnetic systems~\cite{nagaosa013}; classifying defects in liquid crystals~\cite{volume7}; classifying vacuum states in non--Abelian gauge theories~\cite{srednicki2007}; the theory of general relativity in $2+1$ dimensions~\cite{carlip2003}; and the theory of topological insulators in condensed matter physics~\cite{hasan2010}.
\par
This work applies topological methods to electromagnetic materials, predicting mode characteristics that are independent of the detailed inhomogeneity of the material. The results are rather different from theories such as transformation optics~\cite{greenleaf2003,pendry2006,leonhardt2006}, where the function of a device depends on an accurate implementation of the material tensors across space.  Topological results are by definition insensitive to the local details of the material, and have already been shown to govern the number of interface states between adjoining materials~\cite{haldane2008,wang2009,khanikaev2013,lu2014}.   The theory has been developed for both continuous and periodic media~\cite{davoyan2013,silveirinha2016,silveirinha2018}, and has been connected to the properties of the Dirac equation~\cite{horsley2018,mechelen2019}, which also plays a role in this work.  Because these trapped interface states can be predicted using topological methods, they are rather robust to the details of the interface, and have been experimentally observed to propagate past extreme obstacles without backscattering~\cite{wang2009}.
\par
It is unusual for a confined mode to be insensitive to local variations in a material.  A typical electromagnetic mode can be understood as arising from constructive interference between counter propagating waves.  Any change in the refractive index will change the phase of each component wave and thus change the dispersion of the mode.  For example, the dispersion of a guided mode in a dielectric cylinder depends strongly on the size of the cylinder and the distribution of the permittivity~\cite{kurtz1969}.  By contrast, here we find a large class of confined electromagnetic modes that are insensitive to the local inhomogeneity of the material parameters.  The existence of these modes only depends on the behaviour of the material parameters at infinity (or rather, large distances from the inhomogeneity).  For example, we find a family of media with modes that have a dispersion relation that is invariant to local changes to the material.
\par
To find these modes we make use of an analogy between Maxwell's equations in inhomogeneous media and the Dirac equation~\cite{barnett2014,horsley2018}.  To understand this analogy consider the Dirac equation in two dimensions, for a particle of mass $m$ and energy $\mathcal{E}$ in a gauge field $\boldsymbol{A}=A_x\hat{\boldsymbol{x}}+A_y\hat{\boldsymbol{y}}$
\begin{equation}
	\left(\begin{matrix}0&\mathcal{D}\\\mathcal{D}^{\dagger}&0\end{matrix}\right)\left(\begin{matrix}\psi_{+}\\\psi_{-}\end{matrix}\right)=\left(\begin{matrix}\mathcal{E}-m&0\\0&\mathcal{E}+m\end{matrix}\right)\left(\begin{matrix}\psi_{+}\\\psi_{-}\end{matrix}\right)\label{eq:dirac2d}
\end{equation}
where $\mathcal{D}=-2{\rm i}\,\partial/\partial z-A_{+}$, $\psi_{+,-}$ are the wavefunctions for the two spin components, $z=x+{\rm i}y$, and $A_{+}=A_x-{\rm i}A_{y}$.  Besides being a limiting case of the relativistic description of electrons, this equation appears as an effective description in planar optics, notably in deformed honeycomb lattices~\cite{rechtsman2013,mei2012} and gyrotropic media~\cite{horsley2018}.  Indeed, Maxwell's equations for fields of a fixed frequency $\omega$ can be written in a \emph{similar} form if the electric $\boldsymbol{E}$ and magnetic $\boldsymbol{H}$ fields are combined into a single six--vector~\cite{horsley2018}, 
\begin{equation}
	\left(\begin{matrix}
	\boldsymbol{0}&\boldsymbol{\mathcal{D}}\\
	\boldsymbol{\mathcal{D}}^{\dagger}&\boldsymbol{0}
	\end{matrix}\right)\left(\begin{matrix}\boldsymbol{E}\\\eta_0\boldsymbol{H}\end{matrix}\right)=-k_0\left(\begin{matrix}
	\boldsymbol{\epsilon}&\boldsymbol{0}\\
	\boldsymbol{0}&\boldsymbol{\mu}\end{matrix}\right)\left(\begin{matrix}\boldsymbol{E}\\\eta_0\boldsymbol{H}\end{matrix}\right)\label{eq:maxwell6}
\end{equation}
where $\eta_0=\sqrt{\mu_0/\epsilon_0}$ is the impedance of free space, $k_0=\omega/c$ is the free space wavenumber, and $\boldsymbol{\epsilon}$, $\boldsymbol{\mu}$ and $\boldsymbol{\xi}$ are respectively the permittivity, permeability and bianisotropy tensors for a lossless medium.  In this case the differential operator is given by $\boldsymbol{\mathcal{D}}=-{\rm i}\boldsymbol{\nabla}\times+k_0\boldsymbol{\xi}$.  A comparison between equations (\ref{eq:dirac2d}) and (\ref{eq:maxwell6}) shows that, broadly speaking the electric and magnetic fields play the role of the two spin components in an effective Dirac equation: the bianisotropy plays the role of the gauge field, the `energy' is given by $-k_0(\boldsymbol{\mu}+\boldsymbol{\epsilon})/2$, and the `mass' by $k_0(\boldsymbol{\epsilon}-\boldsymbol{\mu})/2$.  Although the vector nature of the wavefunction components make this analogy incomplete, for one dimensional variations we can make it exact. 
\par
Now for the role of topology: if both the energy $\mathcal{E}$ and mass $m$ are zero in the Dirac equation (\ref{eq:dirac2d}), the two spin components $\psi_{\pm}$ become decoupled, satisfying
\begin{align}
	\mathcal{D}\psi_{-}&=0\label{eq:ker1}\\
	\mathcal{D}^{\dagger}\psi_{+}&=0\label{eq:ker2}.
\end{align}
The difference in the number of solutions to (\ref{eq:ker1}), $N={\rm dim}[{\rm ker}\,\mathcal{D}]$ and the solutions to (\ref{eq:ker2}), $\bar{N}={\rm dim}[{\rm ker}\,\mathcal{D}^{\dagger}]$ is governed by a rather deep and far--ranging result in topology known as the Atiyah--Singer index theorem~\cite{atiyah1963,rosenberg1997}.  This theorem is an extreme generalization of the Gauss--Bonnet theorem~\cite{rosenberg1997}, and has been connected to the aforementioned work on interface states between periodic media~\cite{volovik2003,niemi1984}.  In general it states that
\begin{align}
	{\rm index}[\mathcal{D}]&=N-\bar{N}\nonumber\\
	&=\int_{M}\hat{A}(M)\wedge{\rm ch}(V)\label{eq:dirac-index}
\end{align}
where the integration is taken over the manifold $M$, `$\wedge$' is the exterior product, $\hat{A}(M)$ is the `A--hat genus', depending on the curvature of the space, and ${\rm ch}(V)$ is the `Chern character' depending on the curvature of the gauge field~\cite{wassermann2010,getzler1986,mostafazadeh1994}.  To put this theorem in physical terms, the difference in the number of solutions $N-\bar{N}$ cannot be altered though any continuous change of the system parameters.  The implications of this index theorem are well known for the true Dirac equation, but do not seem to have been considered seriously in electromagnetism.  Are electromagnetic modes also controlled by this theorem?  This seems a natural question to ask, given the close similarity between (\ref{eq:dirac2d}) and (\ref{eq:maxwell6}).
%
%
\begin{figure}[h!]
    \centering
    \includegraphics[width=\columnwidth]{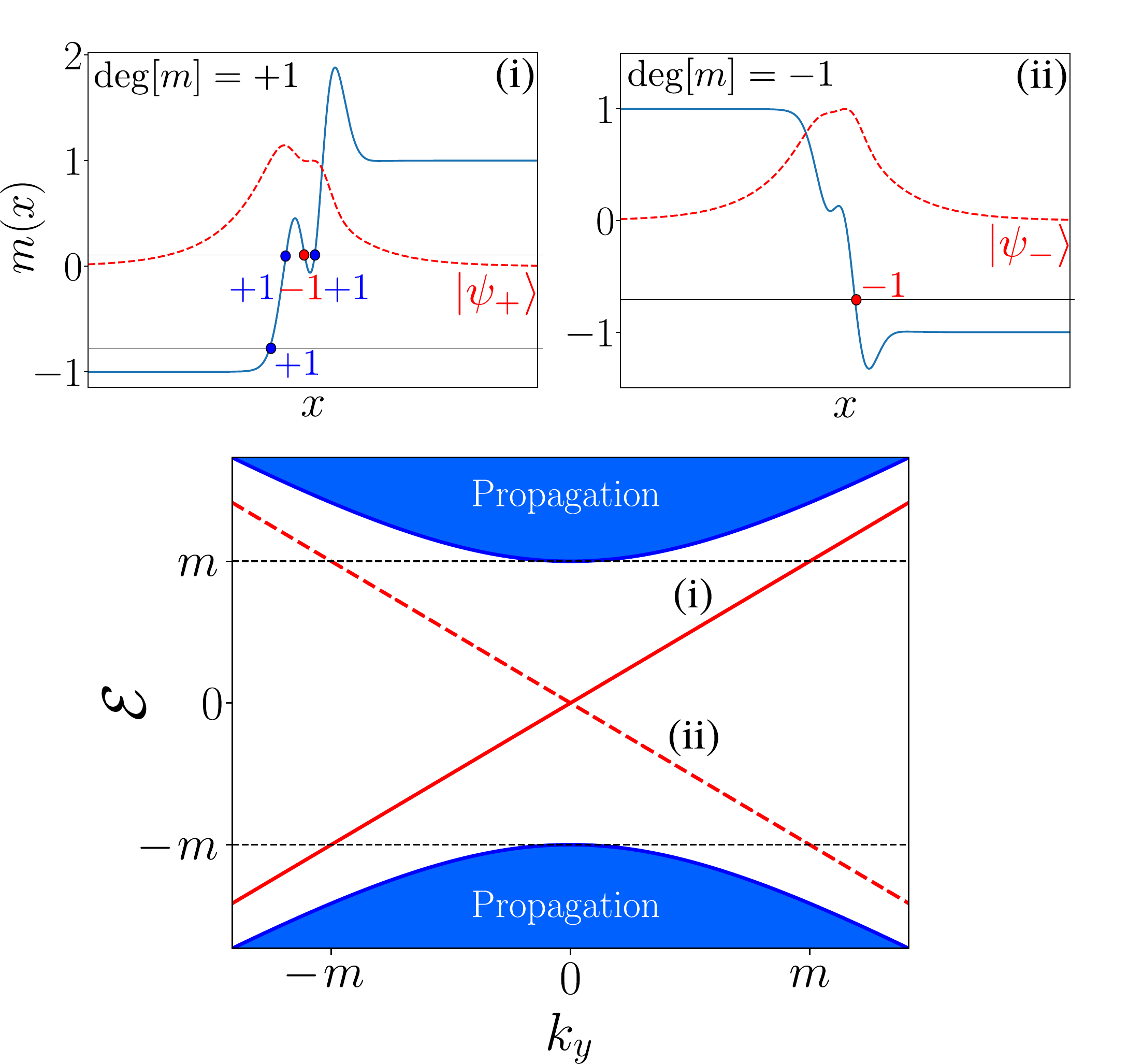}
    \caption{Jackiw--Rebbi modes of the Dirac equation (\ref{eq:2d_dirac}) with position dependent mass.  Different arbitrarily chosen mass distributions are shown as solid blue curves in panels (i) and (ii), plus mode profiles $\psi_{\pm}$ (red dashed lines, computed from (\ref{eq:edge_states})).  The mass profiles can be classified in terms of their topological degree (see Eq. (\ref{eq:degree})), the calculation of which is performed through summing intersection points along the thin horizontal lines as indicated in (i--ii).  The degree is a topological invariant, independent of the choice of horizontal line (assuming the mass ultimately diverges at infinity, without changing sign).  The lower plot sketches the dispersion relation for the Dirac equation (\ref{eq:2d_dirac}), where for $\mathcal{E}>m$ (shaded blue) we have propagation in the region of space where $m\sim{\rm const}.$.  The red lines crossing the `mass gap' $\mathcal{E}\in[-m,m]$ indicate the dispersion of the two modes shown in the upper panels.}
    \label{fig:dirac_dispersion}
\end{figure}
%
%
\subsection{Jackiw--Rebbi modes\label{sec:jr_modes}}
\par
Before considering the electromagnetic case, we review a one dimensional example of the Dirac equation where the allowed modes depend on the behaviour of the system at infinity.  The modes given in this section are the so--called Jackiw--Rebbi modes~\cite{jackiw1976}.
\par
The two--dimensional Dirac equation (\ref{eq:dirac2d}) with zero magnetic field takes the following form
\begin{equation}
    \left[-{\rm i}\sigma_{x}\partial_x-{\rm i}\sigma_{y}\partial_{y}+m(x)\sigma_{z}\right]|\psi\rangle=\mathcal{E}|\psi\rangle\label{eq:2d_dirac}
\end{equation}
where the particle mass now depends on position, and $|\psi\rangle$ is the two component wavefunction.  Assuming translational symmetry along $y$ ($-{\rm i}\partial_{y}\to k_{y}$), and writing the wave in terms of the eigenfunctions of $\sigma_{y}$ with eigenvalue $\pm1$
\begin{equation}
	|\psi\rangle=\psi_{+}a_{y,+}+\psi_{-}a_{y,-}
\end{equation}
where $a_{y,\pm}=(1,\pm {\rm i})^{T}$, the Dirac equation (\ref{eq:2d_dirac}) can be reduced to the form
\begin{equation}
	\left(\begin{matrix}0&\mathcal{D}\\\mathcal{D}^{\dagger}&0\end{matrix}\right)\left(\begin{matrix}\psi_{+}\\\psi_{-}\end{matrix}\right)=\left(\begin{matrix}\mathcal{E}-k_y&0\\0&\mathcal{E}+k_y\end{matrix}\right)\left(\begin{matrix}\psi_{+}\\\psi_{-}\end{matrix}\right)\label{eq:dirac_edge}
\end{equation}
where $\mathcal{D}=-\partial_x+m(x)$.  From the above pair of equations (\ref{eq:dirac_edge}) we see that the difference in the number $N$ of solutions to $\mathcal{D}\psi_{-}=0$ and the number $\bar{N}$ of solutions to $\mathcal{D}^{\dagger}\psi_{+}=0$ is the difference in the number of modes with $\mathcal{E}=-k_y$ and those with $\mathcal{E}=k_y$.  For positive energy this is the difference in the number of modes propagating down or up the $y$ axis.
\par
As discussed above, the difference $N-\bar{N}$ is fixed by the behaviour of $m(x)$ at infinity.  We do not need the sophisticated mathematics of (\ref{eq:dirac-index}) to see this: it is immediately evident from the solutions to (\ref{eq:dirac_edge}),
\begin{equation}
	\psi_{\pm}=\exp\left(\mp\int_{0}^{x}m(x')dx'\right)\qquad(\mathcal{E}=\pm k_y)\label{eq:edge_states}
\end{equation}
The solutions (\ref{eq:edge_states}) are the well--known Jackiw--Rebbi modes~\cite{jackiw1976} that occur between regions where the mass has a different sign.  If $m(x)$ has the same sign at $x=+\infty$ and $x=-\infty$ then neither of the states (\ref{eq:edge_states}) is normalizable and $N=\bar{N}=0$.  Conversely if $m(x)$ takes a different sign at these two limits then $N-\bar{N}$ equals $\pm1$, for the respective cases of negative and positive $m$ at $+\infty$.  A compact way of writing this result is in terms of the degree of the mapping ${\rm deg}[m(x)]$~\cite{outerelo2009}
\begin{equation}
    {\rm index}[\mathcal{D}]=N-\bar{N}-=-{\rm deg}[m(x)]=\pm1,0,\label{eq:1dindex}
\end{equation}
where the degree is defined as
\begin{equation}
    {\rm deg}[m(x)]=\sum_{x_{a}\in m^{-1}(a)}{\rm sign}[m'(x_a)].\label{eq:degree}
\end{equation}
In this one dimensional case the degree is the topological invariant appearing on the right hand side of the index theorem (\ref{eq:dirac-index}).  Examples are shown in panels (i) and (ii) of Fig.~\ref{fig:dirac_dispersion}.  As can be established from an examination of Fig.~\ref{fig:dirac_dispersion}, strictly speaking the mass $m(x)$ should diverge at infinity for the degree ${\rm deg}[m]$ to be well defined~\cite{witten1982}.  In practice however, the modes we predict do not depend on this restriction.
%
%
\subsection{Electromagnetic modes in stratified media}
\par
In electromagnetic terms a Jackiw--Rebbi mode is a bound mode in a stratified medium where the dispersion relation connecting the frequency and wave--vector is insensitive to the specific spatial distribution of the material parameters.  We now tackle the case of generic stratified electromagnetic materials, illustrating why confined modes usually have a dispersion relation sensitive to the precise distribution of material parameters.
\par
For generic stratified media inhomogeneous along $x$, Eq. (\ref{eq:maxwell6}) reduces to
\begin{align}
	{\rm i}\,\Gamma_{1}\frac{d}{dx}\left(\begin{matrix}\boldsymbol{E}\\\eta_0\boldsymbol{H}\end{matrix}\right)&=k_0\left(\begin{matrix}\boldsymbol{\epsilon}&\bar{\boldsymbol{\xi}}\\\bar{\boldsymbol{\xi}}^{\dagger}&\boldsymbol{\mu}\end{matrix}\right)\left(\begin{matrix}\boldsymbol{E}\\\eta_0\boldsymbol{H}\end{matrix}\right)=k_0\chi\left(\begin{matrix}\boldsymbol{E}\\\eta_0\boldsymbol{H}\end{matrix}\right)\label{eq:1dmaxwell6}
\end{align}
where we have introduced a set of matrices $\Gamma_{j}$ to represent the curl operator
\begin{equation}
	\Gamma_{j}=\left(\begin{matrix}\boldsymbol{0}&\boldsymbol{L}_{j}\\-\boldsymbol{L}_{j}&\boldsymbol{0}\end{matrix}\right)
\end{equation}
with the three angular momentum matrices given by
\begin{equation}
	\boldsymbol{L}_{1}=\left(\begin{matrix}0&0&0\\0&0&-1\\0&1&0\end{matrix}\right),\boldsymbol{L}_{2}=\left(\begin{matrix}0&0&1\\0&0&0\\-1&0&0\end{matrix}\right),\boldsymbol{L}_{3}=\left(\begin{matrix}0&-1&0\\1&0&0\\0&0&0\end{matrix}\right).
\end{equation}
The propagation constants along $y$ and $z$ are $k_y$ and $k_z$ respectively, and we have combined the terms arizing from this propagation into an effective bianisotropy tensor $\bar{\boldsymbol{\xi}}$
\begin{equation}
	\bar{\boldsymbol{\xi}}=\boldsymbol{\xi}+\frac{k_y}{k_0}\boldsymbol{\Gamma}_{2}+\frac{k_{z}}{k_0}\boldsymbol{\Gamma}_{3}
\end{equation}
The problem with comparing equation (\ref{eq:1dmaxwell6}) to the Dirac equation is that $\Gamma_{1}$ is not an element of a Clifford algebra.  This is due to the transverse nature of the electromagnetic field, where any field vector with $\boldsymbol{E}$ and $\boldsymbol{H}$ pointing along $\hat{\boldsymbol{x}}$ is reduced to zero by $\Gamma_{1}$.  For planar media we can sidestep this difficulty through solving (\ref{eq:1dmaxwell6}) for the field components $E_x$ and $H_x$, finding that
\begin{equation}
\left(\begin{matrix}E_{x}\\\eta_{0}H_{x}\end{matrix}\right)=-\chi_{xx}^{-1}\chi_{xp}\left(\begin{matrix}\boldsymbol{E}_{\parallel}\\\eta_{0}\boldsymbol{H}_{\parallel}\end{matrix}\right),\label{eq:elimination}
\end{equation}
where $\chi_{xx}$ is the $2\times2$ matrix with elements $\epsilon_{xx}$, $\bar{\xi}_{xx}$, $\bar{\xi}_{xx}^{\star}$, and $\mu_{xx}$ and $\chi_{xp}$ is a $2\times4$ matrix with elements $\boldsymbol{\epsilon_{x\parallel}}$, $\bar{\boldsymbol{\xi}}_{x\parallel}$ etc., and a subscript $\parallel$ indicates components in the $y$--$z$ plane.  Using the result (\ref{eq:elimination}) to eliminate these $\hat{\boldsymbol{x}}$ field components,  Eq. (\ref{eq:1dmaxwell6}) reduces to
\begin{equation}
\gamma_{1}\frac{d}{dx}|\psi\rangle=-{\rm i}k_0\gamma_{1}\gamma_{2}\left[\chi_{pp}-\chi_{px}\chi_{xx}^{-1}\chi_{xp}\right]|\psi\rangle\label{eq:maxwell_reduced}
\end{equation}
where $|\psi\rangle=(\boldsymbol{E}_{\parallel},\eta_{0}\boldsymbol{H}_{\parallel})^{\rm T}$, and the $4\times4$ $\gamma_{j}$ matrices are of the usual Dirac form
\begin{equation}
	\gamma_{0}=\left(\begin{matrix}\boldsymbol{1}_{2}&0\\0&-\boldsymbol{1}_{2}\end{matrix}\right),\gamma_{j}=\left(\begin{matrix}0&\sigma_{j}\\-\sigma_{j}&0\end{matrix}\right),\gamma_{5}=\left(\begin{matrix}0&\boldsymbol{1}_{2}\\\boldsymbol{1}_{2}&0\end{matrix}\right).
\end{equation}
In a planar geometry we can thus reduce Maxwell's equations to a form (\ref{eq:maxwell_reduced}) that is analogous to the four component Dirac equation, with the material parameters corresponding to generally rather complicated contributions to the Hamiltonian.  One advantage of recasting Maxwell's equations into this Dirac--like form is that it is immediately evident that the solution to Eq. (\ref{eq:maxwell_reduced}) can be written in terms of a path ordered exponential
\begin{equation}
	|\psi\rangle={\rm P}\left[{\rm e}^{-{\rm i}k_0\int_{0}^{x}\gamma_{2}\left(\chi_{pp}-\chi_{px}\chi_{xx}^{-1}\chi_{xp}\right)dx'}\right]|\psi_0\rangle\label{eq:general_solution}
\end{equation}
where $|\psi_{0}\rangle$ is the form of the in--plane electromagnetic field at $x=0$.  Importantly this is the general solution to Maxwell's equations in a layered material.
\par
The result given in Eq. (\ref{eq:general_solution}) already resembles the Jackiw--Rebbi mode (\ref{eq:edge_states}).  However they are crucially not the same, and this reveals why the dispersion of confined electromagnetic modes is almost always sensitive to the precise distribution of the material parameters.  Firstly, the exponent of (\ref{eq:general_solution}) is not generally a Hermitian operator.  This means that the exponent can be either real or complex valued.  Secondly, the path ordering is necessary because the basis vectors of the matrix in the exponent (i.e. the polarization basis) will change with position, continually rotating $|\psi_{0}\rangle$ as we move along the $x$ axis.  The combination of these two properties means that an arbitrary choice of $|\psi_{0}\rangle$ in Eq. (\ref{eq:general_solution}) will most often either be propagating or divergent at infinity, rather than tending to zero as the mode (\ref{eq:edge_states}) does.  To find the bound modes in a particular material profile one must carefully choose $k_y$, $k_z$, and $k_0$ such that the field amplitude vanishes asymptotically.  This careful choice is of course the dispersion relation.  From this perspective, electromagnetic Jackiw--Rebbi modes are those special families of material parameters where the path ordering can be dropped from (\ref{eq:general_solution}), and where the exponent is real valued.  Using our analogy with the Dirac equation, we shall now show examples of such confined modes, the existence of which can also be understood in terms of the topological invariant (\ref{eq:degree}).
%
%
\subsection{Examples of electromagnetic Jackiw--Rebbi modes\label{sec:emjr}}
\par
To keep the discussion simple we do not work in terms of our general solution (\ref{eq:general_solution}), but rather specialize to a particular case of the optical Dirac equation (\ref{eq:maxwell_reduced}).  Assuming zero propagation constant along $z$, and modes that are either $E_z$ or $H_z$ polarized waves (i.e. have either their electric, or magnetic fields pointing only along the $\hat{\boldsymbol{z}}$ axis), we are restricted to the following form of the material tensors
\begin{equation}
    \boldsymbol{\epsilon}=\left(\begin{matrix}\boldsymbol{\epsilon}_{\parallel}&\boldsymbol{0}\\\boldsymbol{0}&\epsilon_{zz}\end{matrix}\right),\;\boldsymbol{\mu}=\left(\begin{matrix}\boldsymbol{\mu}_{\parallel}&\boldsymbol{0}\\\boldsymbol{0}&\mu_{zz}\end{matrix}\right),\;\boldsymbol{\xi}=\left(\begin{matrix}\boldsymbol{0}&\boldsymbol{v}^{T}\\\boldsymbol{w}&0\end{matrix}\right)\label{eq:material_params}.
\end{equation}
where $\boldsymbol{v}$ and $\boldsymbol{w}$ are two element complex vectors, i.e. $\boldsymbol{v}=(v_{x},v_{y})$.  With these assumptions (\ref{eq:maxwell_reduced}) reduces to a pair of uncoupled two component Dirac--like equations.  For the $E_z$ polarization the equation is given by
\begin{equation}
\left[-{\rm i}\sigma_{x}\left(k_0^{-1}\frac{d}{d x}+{\rm i}\,\alpha_1\right)-\alpha_2\sigma_{y}+m\sigma_{z}\right]|\psi\rangle
=\mathcal{E}|\psi\rangle\label{eq:maxwell_dirac}
\end{equation}
where $|\psi\rangle=(E_{z},\eta_0 H_y)^{\rm T}$, and the complex number $\alpha$ is given by
\begin{equation}
	\alpha=w_{y}+\frac{\mu_{xy}}{\mu_{xx}}\left(\frac{k_y}{k_0}-w_x\right)
\end{equation}
with its real and imaginary parts labelled as $\alpha=\alpha_{1}+{\rm i}\alpha_{2}$.  The `mass' and `energy' in (\ref{eq:maxwell_dirac}) are given by $m=(\beta-\gamma)/2$ and $\mathcal{E}=(\beta+\gamma)/2$, where
\begin{equation}
	\beta=-\frac{{\rm det}[\boldsymbol{\mu}_{\parallel}]}{\mu_{xx}},\;\;\gamma=-\epsilon_{zz}+\frac{\left|\frac{k_y}{k_0}-w_x\right|^{2}}{\mu_{xx}}.\label{eq:maxwell_m_E}
\end{equation}
A similar formula to the above Dirac--like equation (\ref{eq:maxwell_dirac}) holds for the $H_z$ polarization with permittivity and permeability interchanged, and $\boldsymbol{w}\to\boldsymbol{v}$.  It should be emphasized that the meaning of `energy' and `mass' are here given in terms of material parameters through equation (\ref{eq:maxwell_m_E}), which is conceptually similar to the treatment given in~\cite{horsley2018}.  Despite this difference in interpretation from the true Dirac equation, we can apply the same index theorem summarized in (\ref{eq:edge_states}--\ref{eq:degree}) to identify the existence of `topological' modes within electromagnetic materials, exactly as we did for the true Dirac equation.\\
%
%
\begin{figure}[h!]
\includegraphics[width=\columnwidth]{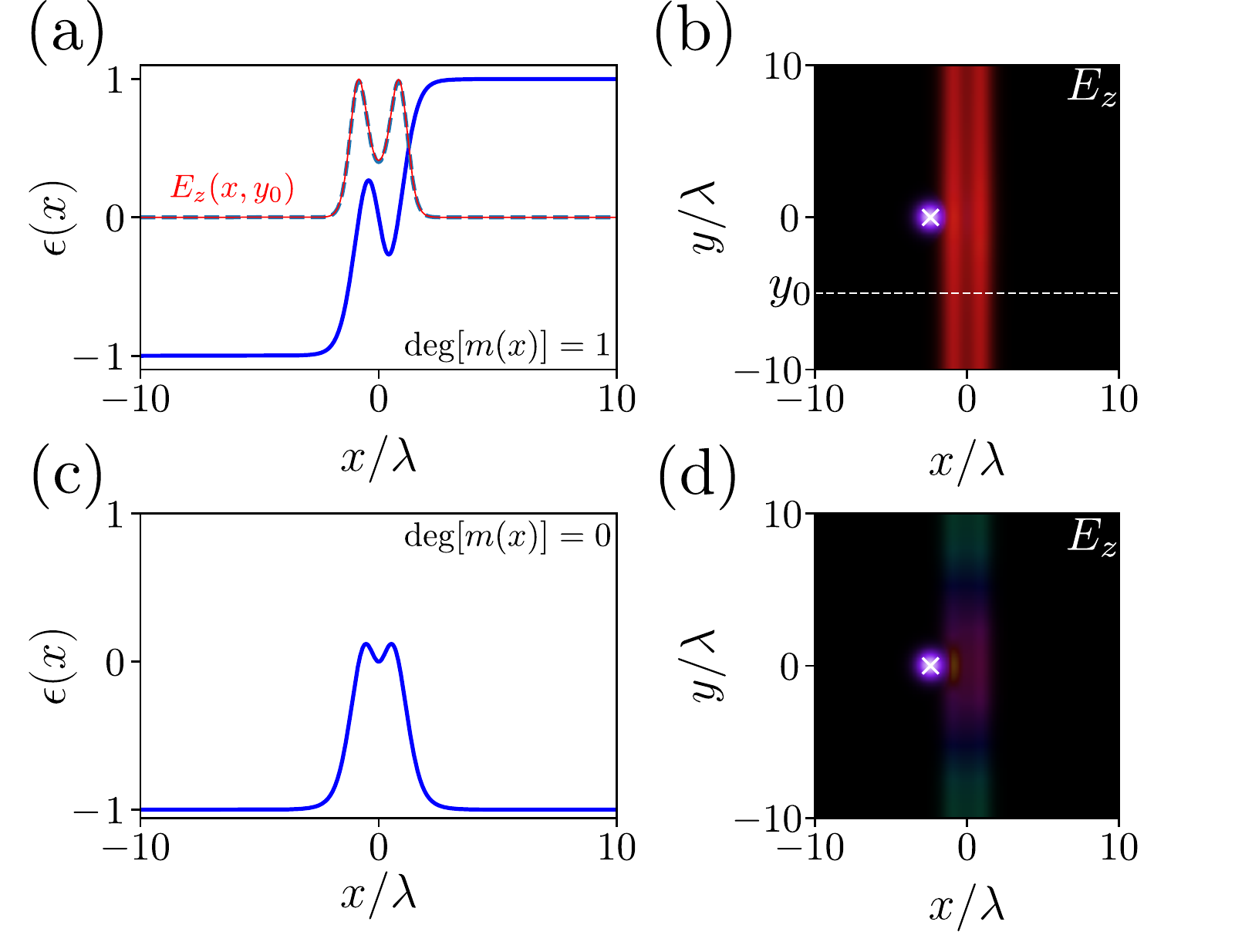}
\caption{A numerical illustration of the Jackiw--Rebbi mode in an isotropic medium where $\epsilon=-\mu=m(x)$.  The blue curves in panels (a) and (c) show two different arbitrarily chosen permittivity profiles with degree $+1$ and $0$ respectively.  According to Eq. (\ref{eq:1dindex}) a confined mode should exist for case (a) but not (c).    Panels (b) and (d) show numerical simulations (COMSOL Multiphysics~\cite{comsol}; note periodic boundary conditions applied along $y$) for cases (a) and (c) respectively, where a line current source of frequency $k_0=2\pi/\lambda$ is positioned at the white cross. The electric field $E_z$ along the line $y=y_0$ is plotted as the red line in panel (a) on top of which (blue dashed line) is plotted the analytical solution (\ref{eq:edge_states}). \label{fig:isotropic_figure}}
\end{figure}
%
%
\paragraph{Isotropic media:}
For the simplest case of a vanishing propagation constant, $k_y=0$, and isotropic materials ($\boldsymbol{\mu}=\mu\boldsymbol{1}$, $\boldsymbol{\epsilon}=\epsilon\boldsymbol{1}$ and $\boldsymbol{\xi}=\boldsymbol{0}$), our Dirac--like equation reproduces the recent findings of Shen et al.~\cite{shen2014}, where it was noticed that planar isotropic media could be understood in terms of a two component Dirac equation.  We give a different viewpoint here, emphasizing the indifference of a bound mode to the details of the inhomogeneity.  For these parameters our general Dirac equation (\ref{eq:maxwell_dirac}) reduces to a simple form where $\alpha=0$ and the mass and energy are given in terms of the difference and the average of the permittivity and permeability, as we anticipated in the general case of equation (\ref{eq:maxwell6})
\begin{align}
	m&=-\frac{1}{2}\left(\mu-\epsilon\right)\nonumber\\
	\mathcal{E}&=-\frac{1}{2}\left(\mu+\epsilon\right).
\end{align}
If the magnetic and electric responses vary in space, but are such that $\epsilon(x)=-\mu(x)=m(x)$ then the solutions to Maxwell's equations become equivalent to the zero energy modes of the Dirac equation (\ref{eq:2d_dirac}), with a position dependent mass $m=m(x)$.  In electromagnetic terms such a medium would not be expected to support any bound modes because  the refractive index $n=\sqrt{\epsilon\mu}={\rm i}m(x)$ is purely imaginary everywhere.  This agrees with the Dirac picture sketched in the lower panel of Fig.\ref{fig:dirac_dispersion}, where zero energy lies in the centre of the energy gap, and no propagation is possible.  However, as we established in section~\ref{sec:jr_modes} (see Eq. (\ref{eq:dirac_edge}--\ref{eq:degree})) there can be modes in such a system.  Their number is again governed by the degree of the function $m(x)$ (see Eq. (\ref{eq:1dindex})).  Therefore an inhomogeneous medium where the permittivity and permeability have equal magnitude and opposite sign supports a single $k_y=0$ mode if the permittivity and permeability have different signs at $+\infty$ and $-\infty$.  The details of the interface are irrelevant.  This is demonstrated numerically in Fig.~\ref{fig:isotropic_figure}.
%
%
\begin{figure}[h!]
\includegraphics[width=\columnwidth]{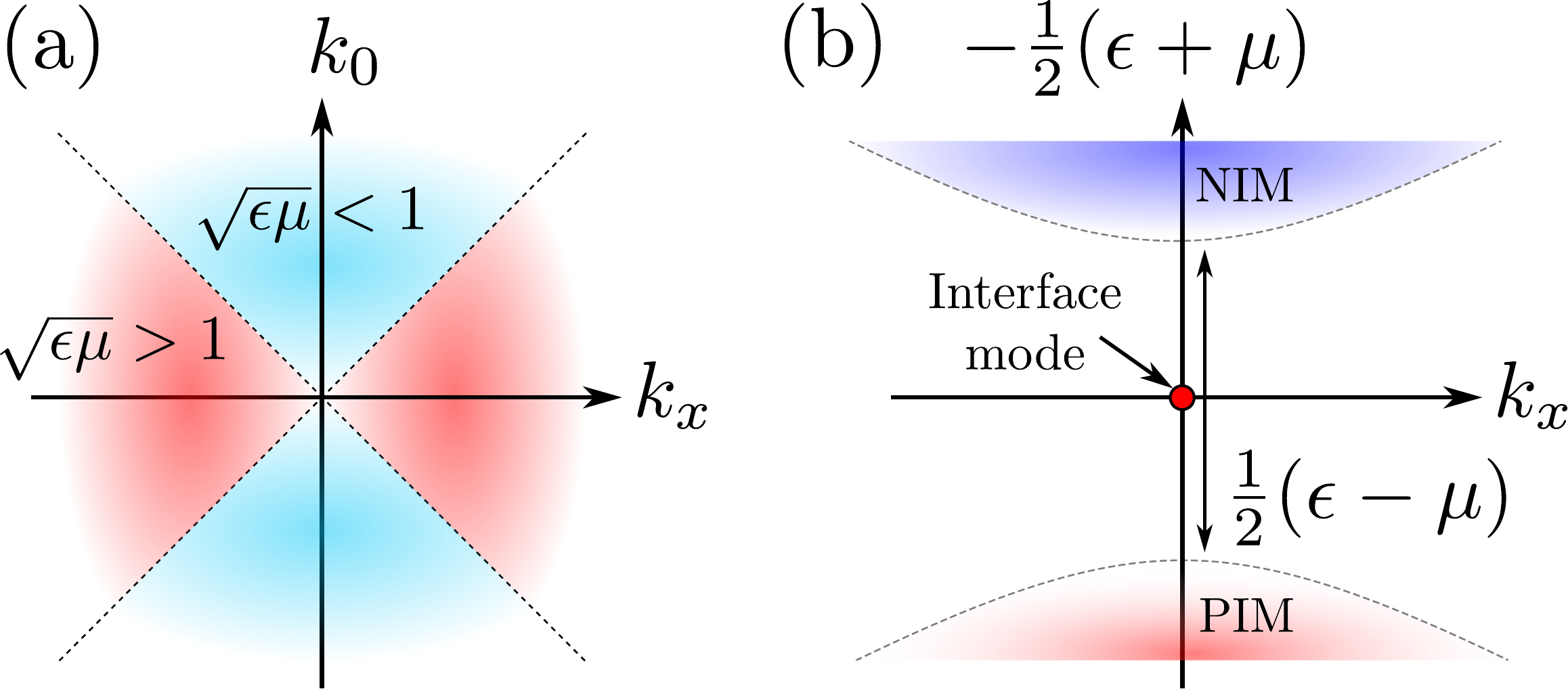}
\caption{Maxwell's equations in an isotropic material can be understood as a special case of the Dirac equation.  (a) The effect of a material is usually understood in terms of the refractive index $n=\sqrt{\epsilon\mu}$, via the dispersion relation $k_x^2=n^2 k_0^2$.  A refractive index greater than unity moves the dispersion cone into the red region, and one of less than unity moves the dispersion into the blue region. (b) An alternative interpretation is to write the dispersion as $(k_x/k_0)^2=\mathcal{E}^2-m^2$, as given in Eq. (\ref{eq:isotropic_dispersion}).  This shows that the region of disallowed propagation $|\epsilon+\mu|<|\epsilon-\mu|$ is equivalent to the `mass gap' shown in the lower panel of Fig.~\ref{fig:dirac_dispersion}, in this case separating regions of positive (PIM) and negative (NIM) index media.  Accordingly there is a Jackiw--Rebbi like mode at $\mathcal{E}=-(\mu+\epsilon)/2=0$, the existence of which depends only on the zero energy condition and the properties of $\mu$ and $\epsilon$ at infinity. \label{fig:isotropic_figureb}}
\end{figure}
\par
This analogy with the Dirac equation also reveals a rather unusual but informative way to understand the dispersion of electromagnetic waves in isotropic materials, which is worth commenting on.  For the case of propagation along $x$ through a homogeneous medium the dispersion relation derived from (\ref{eq:maxwell_dirac}) is given by the expected
\begin{align}
	\left(\frac{k_x}{k_0}\right)^{2}=\mathcal{E}^{2}-m^{2}&=\left(\frac{\epsilon+\mu}{2}\right)^{2}-\left(\frac{\epsilon-\mu}{2}\right)^{2}\nonumber\\
	&=\epsilon\mu.\label{eq:isotropic_dispersion}
\end{align}
However this way of writing the equation reveals an interpretation in terms of an `energy' and a `mass', with propagation only possible when the sum of permittivity and permeability is greater in magnitude than their difference.  This shift in interpretation is sketched in Fig.~\ref{fig:isotropic_figureb}.  If we imagine a family of materials with a fixed difference between $\epsilon$ and $\mu$, this is analogous to a fixed mass in the Dirac equation, and results in a gap in the allowed values of the average $(\epsilon+\mu)/2$, analogous to the `mass gap' shown in the lower panel of Fig.~\ref{fig:dirac_dispersion}.  As an illustrative example consider a non--magnetic material, $\mathcal{E}=-(1+\epsilon)/2$ and $m=-(1-\epsilon)/2$.  The boundary between allowed and forbidden propagation is when $\mathcal{E}=m$, which in this case is when $\epsilon=0$, i.e. the tipping point between dielectric and metallic behaviour.  The gap in the dispersion relation also closes when $m=0$, which equivalently is when $\epsilon=\mu$.  This is the condition for impedance matching, and is when the effect of the material is equivalent to that of a coordinate transformation~\cite{pendry2006}.
\par
As a further comment, note that from (\ref{eq:isotropic_dispersion}) propagating waves are only possible when $\epsilon$ and $\mu$ have the same sign; when both are positive we have positive index, or `right--handed' media, and when both are negative we have `left--handed' or negative index media~\cite{veselago1968,pendry2000}.  The two signs of $\mathcal{E}$ in our analogy thus correspond respectively to positive and negative index media, as indicated by the colouring of the two regions of allowed propagation in Fig.~\ref{fig:isotropic_figureb}b.\\

%
%
\begin{figure}[h!]
\includegraphics[width=\columnwidth]{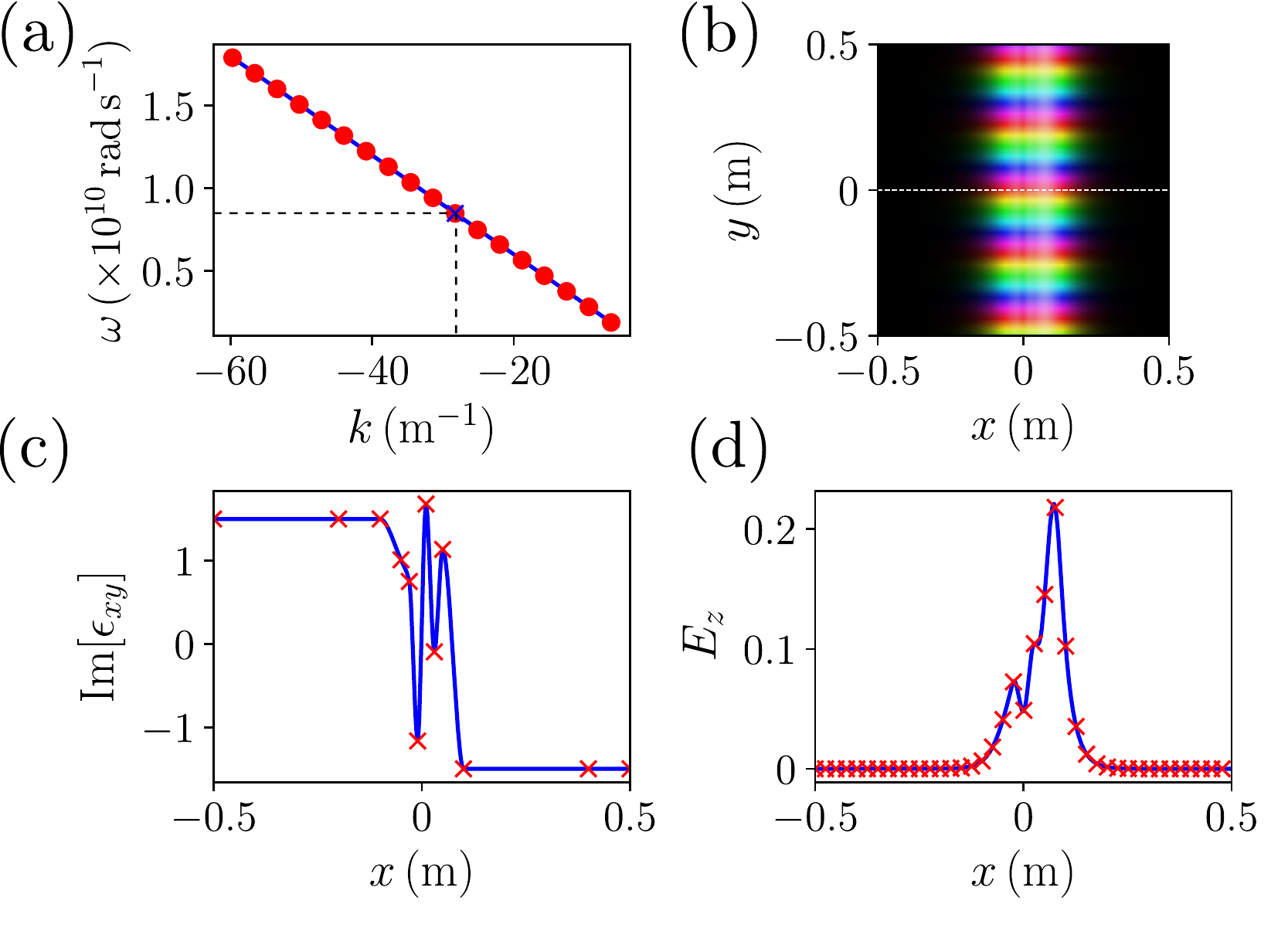}
\caption{One of the confined modes in a medium where the gyrotropy changes sign can be understood as Jackiw--Rebbi mode and has a dispersion relation that is insensitive to how the gyrotropy parameter changes sign. Panel (a) compares the numerically calculated dispersion relation for one such mode (red dots) to the analytical prediction (blue solid line).  The dashed black lines indicate the frequency and wave--vector of the numerically determined mode shown in panel (b).  Panel (c) shows the spatial profile of the gyrotropy parameter, the central region of which was generated using an interpolation of a set of random numbers (red crosses).  Panel (d) compares the numerical mode shape (blue solid line), evaluated along the white dashed line in panel (b), to the analytical prediction (\ref{eq:edge_states})\label{fig:gyrotropic_figure}}
\end{figure}

%
%

\paragraph{Gyrotropic media:\label{sec:gyrotropy}}
A second example of these electromagnetic Jackiw--Rebbi modes is the case of a general complex Hermitian permeability $\boldsymbol{\mu}_{\parallel}$ (zero bianisotropy $\boldsymbol{\xi}=\boldsymbol{0}$).  A complex Hermitian form for either the permeability or the permittivity encodes the physical phenomenon known as gyrotropy~\cite{volume8}, that has already been found to lead to unidirectional propagation of electromagnetic modes~\cite{davoyan2013}, modes that can be counted using topological invariants~\cite{silveirinha2016,horsley2018b,horsley2018}.  As found in~\cite{horsley2018b}, this is intimately connected with the properties of the Dirac equation.  We now show that, without having to compute anything as complicated as a Chern number, very general statements can be made about such materials on the basis of the index theorem summarized in (\ref{eq:edge_states}--\ref{eq:degree}).  
\par
For these gyrotropic media we find that (\ref{eq:maxwell_dirac}) reduces to
\begin{equation}
	\left(\begin{matrix}0&\mathcal{D}\\\mathcal{D}^{\dagger}&0\end{matrix}\right)\left(\begin{matrix}E_z\\\eta_0 H_y\end{matrix}\right)
	=\left(\begin{matrix}{\rm i}\gamma&0\\0&-{\rm i}\beta\end{matrix}\right)\left(\begin{matrix}E_z\\\eta_0 H_y\end{matrix}\right)\label{eq:gyrotropy_index}
\end{equation}
where $\mathcal{D}=k_0^{-1}\partial_x+{\rm i}\alpha_{1}-\alpha_{2}$.  There are solutions to (\ref{eq:gyrotropy_index}) that are in the kernel of either $\mathcal{D}$ or $\mathcal{D}^{\dagger}$
\begin{align}
	\mathcal{D}^{\dagger}E_{z}&=0\qquad\gamma=\frac{1}{\mu_{xx}}\left(\frac{k_y}{k_0}\right)^2-\epsilon_{zz}=0,\;H_y=0\nonumber\\
	\mathcal{D}H_{y}&=0\qquad\beta=-\frac{{\rm det}[\boldsymbol{\mu}_{\parallel}]}{\mu_{xx}}=0,\;E_z=0.\label{eq:gyromodes}
\end{align}
The modes where $\mathcal{D}^{\dagger}E_{z}=0$ and $\mathcal{D}H_{y}=0$ are again of the form (\ref{eq:edge_states}) and are respectively given by
\begin{align}
	E_{z}&=E_{0}\exp\left(-\frac{k_y}{\mu_{xx}}\int_{0}^{x}\left({\rm Im}[\mu_{xy}]+{\rm i}{\rm Re}[\mu_{xy}]\right)dx'\right)\nonumber\\
	H_{y}&=H_{0}\exp\left(\frac{k_y}{\mu_{xx}}\int_{0}^{x}\left({\rm Im}[\mu_{xy}]-{\rm i}{\rm Re}[\mu_{xy}]\right)dx'\right).\label{eq:gyroEzHy}
\end{align}
Whether these states can be normalized is determined by the sign of the imaginary part of $\mu_{yx}$.  A comparison with (\ref{eq:edge_states}--\ref{eq:1dindex}) shows that the index of $\mathcal{D}$ is given by
\begin{equation}
	{\rm index}\left[\mathcal{D}\right]=-{\rm deg}\left[\frac{k_y}{\mu_{xx}}{\rm Im}[\mu_{xy}]\right]\label{eq:gyrotropy_indexa}
\end{equation}
From Eq. (\ref{eq:gyromodes}) we can see that when the index equals $-1$, the propagation constant satisfies the dispersion relation $k_y^{2}=\epsilon_{zz}\mu_{xx}k_0^{2}$ (for simplicity it is assumed that $\epsilon_{zz}\mu_{xx}$ is constant in space).  Eq. (\ref{eq:gyrotropy_indexa}) also shows that this mode corresponds to a degree of $k_y{\rm Im}[\mu_{xy}]/\mu_{xx}$ equal to $+1$.  For $k_y/\mu_{xx}<0$, the imaginary part of $\mu_{xy}$ must thus be positive at $-\infty$ and negative at $+\infty$.  This behaviour is verified in Fig.~\ref{fig:gyrotropic_figure}, where it is shown that the asymptotic behaviour of $\mu_{xy}$ (rather than the local details of the material) determines both the dispersion and the propagation direction of this electromagnetic mode.  Note that, as established in~\cite{horsley2018}, the gyrotropy ${\rm Im}[\mu_{xy}]$ plays the role of the mass in the analogous Jackiw--Rebbi mode (\ref{eq:edge_states}).
\par
Eq. (\ref{eq:gyromodes}) shows that in the second case where ${\rm index}[\mathcal{D}]=1$ we must have $\beta={\rm det}[\boldsymbol{\mu}_{\parallel}]=0$.  This corresponds to the case where one of the eigenvalues of the permeability vanishes.  As shown in Ref.~\cite{horsley2018b}, this is rather special point which can be understood as a zero in the refractive index for propagation in a complex direction.  The mode corresponding to $\mathcal{D}H_y=0$ is unusual because it has no constraint on the magnitude of $k_y$ (i.e. the condition $\beta=0$ does not give rise to a dispersion relation, unlike $\gamma=0$).  This `unconstrained' part of the electromagnetic field was also found in~\cite{horsley2018b}, and has its value determined by the boundary conditions of the system.  For example, if a magnetic mirror is placed anywhere along the $y$--axis this forces $H_y=0$, which eliminates this part of the field.\\
%
%
\begin{figure}[h!]
\includegraphics[width=\columnwidth]{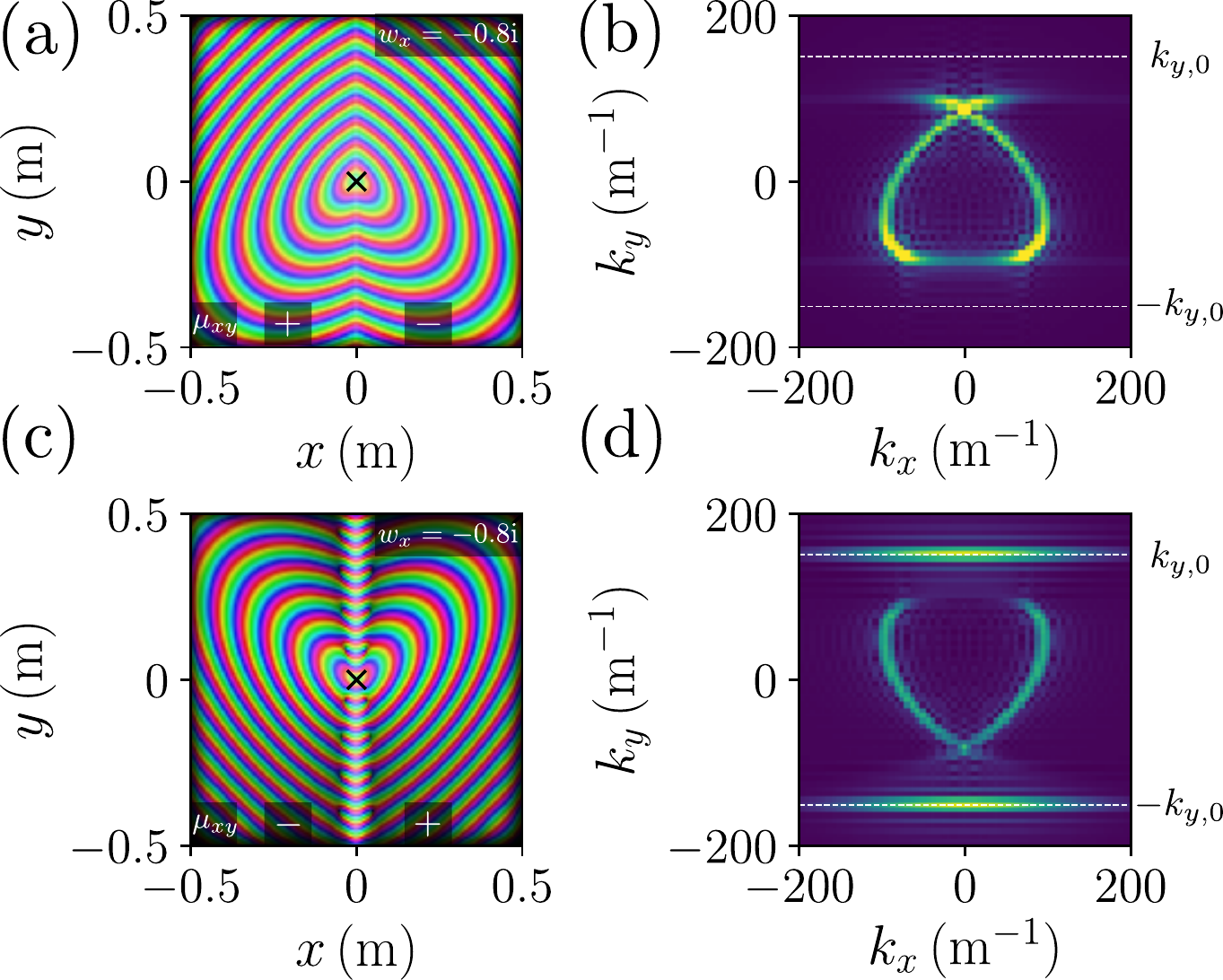}
\caption{Jackiw--Rebbi modes in anisotropic chiral media.  The existence of these modes is governed by the sign of the product of the chirality and the permeability at infinity.  Panel (a) shows the numerical simulation of the field from a point source (black cross), where the index of $\mathcal{D}$ is $+1$, and we do not predict the presence of a localized mode (again we made an arbitrary choice for the spatial profile of $\mu_{xy}$, the asymptotic sign of which is shown at the bottom of each plot).  In panel (b) we   performed the same simulation for an index of $-1$.  Panels (c) and (d) verify the predictions of Eq. (\ref{eq:acindex}) where we plot the numerical Fourier transform~\cite{scipy} of the fields in panels (a) and (c).  The horizontal dashed white lines in panel (d) show the expected dispersion of the confined modes, $k_{y,0}=k_0\sqrt{\mu_{xx}\epsilon_{zz}-\kappa_x^{2}}$.  Note the parameters: $\epsilon_{zz}=\mu_{xx}=1$, $\mu_{xy}\in[-0.5,0.5]$, and $w_y=0$.\label{fig:chiral_anisotropic_figure}}
\end{figure}
%
%
\paragraph{Anisotropic chiral media:}
As a third example the medium exhibits a combination of anisotropy ($\mu_{xy}$ real and non--zero) and chirality (the bianisotropy is imaginary $\boldsymbol{w}={\rm i}\boldsymbol{\kappa}$).  This case doesn't seem to have been considered before, and the optical Dirac equation (\ref{eq:maxwell_dirac}) reduces to the equation (\ref{eq:gyrotropy_index}) as in section~\ref{sec:emjr}b.  In this case the equivalent to the modes (\ref{eq:gyroEzHy}) are given by
\begin{equation}
	E_z=\exp\left\{k_0\int_0^x\left[\frac{\mu_{xy}}{\mu_{xx}}\left(\kappa_x-\frac{{\rm i}k_y}{k_0}\right)- \kappa_{y}\right]dx'\right\}
\end{equation}
and
\begin{equation}
	H_y=\exp\left\{-k_0\int_0^x\left[\frac{\mu_{xy}}{\mu_{xx}}\left(\kappa_x-{\rm i}\frac{k_y}{k_0}\right)- \kappa_{y}\right]dx'\right\}.
\end{equation}
Therefore the difference in the number of solutions to $\mathcal{D}H_{y}=0$ and $\mathcal{D}^{\dagger}E_z=0$, $N-\bar{N}$ is given by
\begin{equation}
	{\rm index}[\mathcal{D}]={\rm deg}\left[\frac{\mu_{xy}}{\mu_{xx}}\kappa_{x}-\kappa_{y}\right].\label{eq:acindex}
\end{equation}
We assume only the permeability component $\mu_{xy}$ changes with position, with everything else constant over space.  If the degree equals $-1$ then, as in the previous example the mode must have the dispersion relation $k_y^{2}=k_0^{2}(\mu_{xx}\epsilon_{zz}-\kappa_x^2)$, although in this case the sign of $k_y$ is not restricted.  Fig.~\ref{fig:chiral_anisotropic_figure} shows a numerical verification of this effect, again for an arbitrary choice of spatially dependent $\mu_{xy}$.  Our formalism predicts slightly unusual materials where a pair of linearly dispersing electromagnetic modes exist, provided that $\kappa_{y}-\mu_{xy}\kappa_{x}/\mu_{xx}]$ increases from a negative value at $-\infty$ to a positive one at $+\infty$, but not the reverse.
\\
%
%
\subsection{Non--Hermitian materials\label{sec:non-hermitian}}
\par
So far our results concern bound modes within lossless media.  In section we shall consider non--Hermitian media where the modes are propagating rather than bound, showing that the topological invariant (\ref{eq:degree}) can also govern the behaviour of propagating electromagnetic waves.  Non--Hermitian systems have recently attracted interest in electromagnetism~\cite{longhi2018}, providing a different route to realise reflectionless~\cite{horsley2015} and invisible~\cite{lin2011} materials.  There are some interesting implications of the above results in non--Hermitian systems.  Analogues of the Jackiw--Rebbi modes exist in media with profiles of loss and gain.  We shall show that the degree (\ref{eq:degree}) controls not the index of the operator, but the character of the modes; either indicating a lack of reflection, coherent perfect absorption or lasing.  This explains some recent findings of Makris and coworkers~\cite{makris2017} who discovered a family of complex profiles that do not exhibit reflection, for an arbitrary amount of disorder. 
\par
Although we could begin the discussion from our earlier general point of view (\ref{eq:maxwell_reduced}), we can make the same point in a simple example.  Take a fixed frequency, TE polarized wave propagating through an isotropic dielectric ($\mu=1$) where the complex dielectric constant $\epsilon(x)$ varies in one spatial direction only.  For propagation along $x$, Maxwell's equations reduce to the one dimensional Helmholtz equation for the wave amplitude $E_{z}$
\begin{equation}
	\left[\frac{d^{2}}{d x^{2}}+k_0^{2}\epsilon(x)\right]E_z=0
\end{equation}
Rather than considering separate field components as we did in the previous section, we now separate the electric field amplitude into its real and imaginary parts $E_z=E_1+{\rm i}E_{2}$.  This scalar equation for the complex variable $E_{z}$ thus becomes a pair of coupled equations for the real variables $E_1$ and $E_2$
\begin{equation}
\left[\frac{d^{2}}{d x^{2}}+k_0^{2}\epsilon_1(x)-{\rm i}k_0^{2}\epsilon_{2}(x)\sigma_{y}\right]|\psi\rangle=0\label{eq:matrix_helmholtz}
\end{equation} 
where $|\psi\rangle=(E_{1},E_{2})^{\rm T}$ and the permittivity was written $\epsilon=\epsilon_{1}+{\rm i}\epsilon_{2}$.  For a certain class of complex permittivity profiles $\epsilon(x)$ the second order operator (\ref{eq:matrix_helmholtz}) can be written as the square of a Dirac operator
\begin{align}
\mathcal{D}^{2}&=\left(-{\rm i}\sigma_{x}\frac{d}{d x}-{\rm i}\sigma_{z}k_0\alpha(x)\right)^{2}\nonumber\\
&=-\frac{d^{2}}{d x^{2}}-k_0^2\alpha^{2}(x)+{\rm i}k_0\sigma_y\alpha'(x)\label{eq:helmholtz_factor}
\end{align}
where $\alpha(x)$ is some real valued function of position.  This idea is similar to that used by Longhi~\cite{longhi2010b}, who identified the transfer matrix formula as an effective Dirac operator.  Equation (\ref{eq:helmholtz_factor}) shows that when the Helmholtz operator equals $\mathcal{D}^{2}$, the permittivity is a complex function of the form
\begin{equation}
	\epsilon(x)=\alpha^{2}(x)+\frac{{\rm i}}{k_0}\alpha'(x).\label{eq:permittivity}
\end{equation}
This form of the permittivity was recently considered in~\cite{makris2017} because they found that, however disordered such a profile, it could support waves that propagate without either backscattering or intensity variation.  Here we see these properties are a consequence of the factorization given in (\ref{eq:helmholtz_factor}), and are another example of Jackiw--Rebbi modes appearing in electromagnetism.
\par
One of the solutions to the above equation is that where $\mathcal{D}|\psi\rangle=0$, which is again the one dimensional Dirac equation for zero energy
\begin{equation}
	\left[-{\rm i}\sigma_{x}\frac{d}{dx}-{\rm i}\sigma_{z}k_0\alpha(x)\right]|\psi\rangle=0.\label{eq:imaginary_mass}
\end{equation}
%
%
\begin{figure}[h!]
\includegraphics[width=\columnwidth]{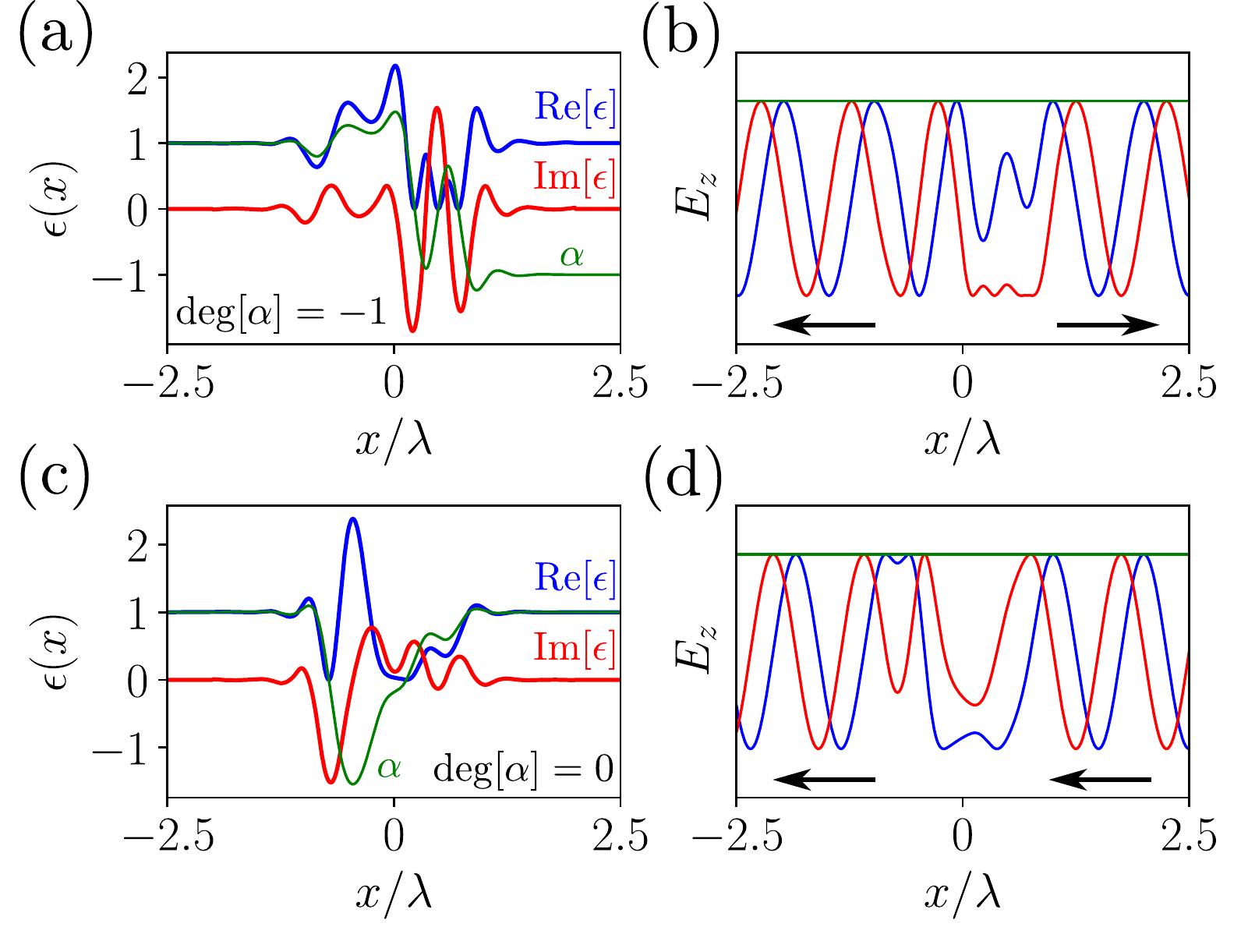}
\caption{For the complex permittivity profiles (\ref{eq:permittivity}), the Helmholtz equation becomes an effective Dirac equation with an imaginary mass $m=-{\rm i}k_0\alpha(x)$.  Two example profiles (central region constructed using an interpolation of random numbers) are given in panels (a) and (c), with the corresponding waves in panels (b) and (d) (real part blue, imaginary part red, and magnitude green).  The wave equation was integrated numerically using functions from the \python Scipy library~\cite{scipy}.   For imaginary mass, the Jackiw--Rebbi modes (\ref{eq:edge_states}) are constant amplitude travelling waves, and the degree of $\alpha$ determines, via (\ref{eq:degree_nh}), whether the mode is a travelling wave, is purely out--going, or purely in--coming.\label{fig:non_hermitian_figure}}
\end{figure}
As we have seen many times, there are solutions (\ref{eq:edge_states}) to this equation not governed by the detailed behaviour of the `mass' $m(x)$, but by it's value at $\pm\infty$.  The difference in this case is that the `mass' is an imaginary quantity $m(x)=-{\rm i}k_0\alpha(x)$ and the two solutions (\ref{eq:edge_states}) to Eq. (\ref{eq:imaginary_mass}) are given by
\begin{equation}
	|\psi\rangle=\frac{1}{2}
	\begin{cases}
	a_{y,+}{\rm e}^{{\rm i}\int_{0}^{x}\alpha(x')dx'}+a_{y,-}{\rm e}^{-{\rm i}\int_{0}^{x}\alpha(x')dx'}\\[10pt]
	{\rm i}a_{y,-}{\rm e}^{-{\rm i}\int_{0}^{x}\alpha(x')dx'}-{\rm i}a_{y,+}{\rm e}^{{\rm i}\int_{0}^{x}\alpha(x')dx'}\label{non-hermitian-solution}
	\end{cases}
\end{equation}
where the linear combinations of (\ref{eq:edge_states}) have been chosen so that the components of $|\psi\rangle$ are real valued.  In the case of an imaginary mass, the degree of ${\rm Im}[m(x)]$ is not related to the kernel of $\mathcal{D}$. This is because the `mass' now only controls the phase of the solution, and whatever it's sign at infinity this is irrelevant to the norm of $|\psi\rangle$.  Nevertheless the degree of $\alpha$ still controls something important about the wave.
\par
Writing the solution (\ref{non-hermitian-solution}) in component form we have
\begin{equation}
	|\psi\rangle=\left(\begin{matrix}\cos\left(\int_0^{x}\alpha(x')dx'\right)\\-\sin\left(\int_0^{x}\alpha(x')dx'\right)\end{matrix}\right),\,\left(\begin{matrix}\sin\left(\int_0^{x}\alpha(x')dx'\right)\\\cos\left(\int_0^{x}\alpha(x')dx'\right)\end{matrix}\right)
\end{equation}
both of which correspond to the same wave, $E_{z}=\exp(-{\rm i}\int_{0}^{x}\alpha(x')dx')$ which is left going is $\alpha$ is positive as $x\to\pm\infty$.  We can thus see that if the degree of $\alpha$ is zero, the material supports a wave of constant amplitude that propagates either to the left or the right, depending on the sign of $\alpha$, without reflection.  Meanwhile if the degree of $\alpha$ is $-1$ the wave is outgoing on both the right and the left hand side of the profile, and the material thus acts as a `laser'.  Finally, for a degree of $+1$ the wave is incoming on the left and the right of the profile and we have so--called coherent perfect absorption (CPA)~\cite{chong2010,longhi2010a}.  This can be summarized as
\begin{equation}
	{\rm deg}[\alpha(x)]=\begin{cases}+1&{\rm (CPA)}\\-1&{\rm (Lasing)}\\0&{\rm (No\; reflection)}\end{cases}\label{eq:degree_nh}
\end{equation}
Fig. \ref{fig:non_hermitian_figure} shows a numerical demonstration of this effect, where two $\alpha$ profiles have been constructed using an interpolation of random numbers, with $|\alpha|\to1$ at infinity.  Although no longer a consequence of the Atiyah--Singer index theorem, the degree of $\alpha(x)$ appearing in the permittivity profiles (\ref{eq:permittivity}) determines something about the wave that is again independent of the detailed behaviour of the material profile.
\subsection{Summary and Conclusions}

In this work we investigated the electromagnetic analogues of the Jackiw--Rebbi modes of the Dirac equation, illustrating that for some families of stratified electromagnetic materials, one can vary the material in an arbitrary fashion without changing the dispersion of one of the bound modes.  The examples considered here show that known modes of both isotropic and gyrotropic media can be understood in this way, and one can also predict new unusual modes such as the example given for anisotropic chiral materials.  In all these cases the existence of the mode can be determined using the same simple topological invariant.
\par
Finally we showed in section~\ref{sec:non-hermitian}, that these applications are not restricted to bound states and one can use the same topological invariant to predict the character of non--Hermitian media via formula (\ref{eq:degree_nh}).  This result showed that the recent discovery of disordered scattering free non--Hermitian media~\cite{makris2017} is actually an instance of a Jackiw--Rebbi mode in electromagnetism.
\par
We also found that for stratified media the Maxwell equations can be written as a four component Dirac equation, from which we can find the general solution as the path ordered product (\ref{eq:general_solution}).  Aside from the examples given here, there seem to be many more interesting applications of this formula.
\acknowledgements
SARH acknowledges useful conversations with Bill Barnes, Tom Philbin, and a series of illuminating lectures from Andrey Shytov.  SARH is funded by the Royal Society and TATA (RPG-2016-186).
%
%
\newpage
\appendix

\bibliography{refs}{}
\bibliographystyle{unsrt}
\end{document}